\documentstyle[12pt,epsfig]{article}
\title{Using Spectral Method as an Approximation for Solving Hyperbolic PDEs}
\author{P. Pedram\thanks{Email: pedram@sbu.ac.ir}, M. Mirzaei and S. S. Gousheh
\\ {\small Department of Physics, Shahid Beheshti University,
Evin, Tehran 19839, Iran}}
\begin{document}
\maketitle \baselineskip 24pt
\begin{abstract}
We demonstrate an application of the spectral method as a numerical
approximation for solving Hyperbolic PDEs. In this method a finite
basis is used for approximating the solutions. In particular, we
demonstrate a set of such solutions for cases which would be
otherwise almost impossible to solve by the more routine methods
such as the Finite Difference Method. Eigenvalue problems are
included in the class of PDEs that are solvable by this method.
Although any complete orthonormal basis can be used, we discuss two
particularly interesting bases: the Fourier basis and the quantum
oscillator eigenfunction basis. We compare and discuss the relative
advantages of each of these two bases. \vspace{5mm}
   \vspace{15mm}
\end{abstract}

\section{Introduction}
Partial differential equations are ubiquitous in science and
industry, since the dynamical laws governing all  physical
phenomena can be usually approximated by a set of partial
differential equations. These include diverse phenomena such as
the dynamics of fluids, gravitational fields, electromagnetic
fields, {\it etc}. In particular, we are more interested in the
applications in quantum cosmology and quantum mechanics.

In some particular physical applications the problem is so
simplified that the resulting PDE is simple enough to be solved
analytically, for example by the method of separation of variables.
However, when a more realistic modeling of the problem is required,
the resulting partial differential equation might become so
complicated that an analytical solution might not be feasible any
more. In such cases, we have to resort to an approximate technique.
These approximate techniques could be either analytic or numeric. An
obvious example of an approximate analytic technique would be the
usual perturbation method, in which the problem is divided into two
segments. The main segment is supposed to be exactly solvable. The
second segment is supposed to modify the solution obtained in the
main segment only very slightly. This modification can be obtained
analytically for any desired degree of accuracy. On the other hand,
the numeric solutions could be either perturbative in nature or
completely numeric. In the first case, as explained above, the main
part is solved analytically. However, the perturbation part is
solved numerically.  Examples of the completely numerical methods
range from the simple Finite Difference Methods (FDM) \cite{FD} to
the more sophisticated Multigrid Method, Collocation Method
\cite{COL1,COL2}, and Finite Element Method (FEM) \cite{EM1,EM2}. In
these methods the configuration space is discretized, and the value
of the solution at each grid point is determined by the values of
its neighboring points. Therefore in these methods the smoothness of
the solution is an important condition to get a reasonable
approximate solution. However, sometimes the solutions are not
smooth. Moreover, we have encountered problems which seem to posses
intrinsic instabilities that we were not able to overcome, even when
we implemented the Courant stability condition
\cite{numerical_recipes} using FDM, or the implicit FDM, or their
combination. The simplest problem we have encountered that seems to
posses both of the aforementioned problems is the following
hyperbolic PDE. This problem appeared in applications of
Robertson-Walker quantum cosmology with zero curvature and is a
particular case of the so called the Wheeler-DeWitt (WD) Eq.\
 \cite{dewitt},
\begin{eqnarray}
\left\{-\frac{\partial^2}{\partial u^2}+\frac{\partial^2}
{\partial v^2}+ u^2- v^2\right\}\psi(u,v)=0. \label{eqwheeler}
\end{eqnarray}
This equation is exactly solvable \cite{wavepacket}. By choosing
an appropriate set of initial conditions, the absolute value
squared of the solution (sometimes called the wave packet) has a
smooth behavior and coincides well with the classical solution
\cite{wavepacket}. However, if we consider the real and imaginary
parts of the solution separately, we can see that each part has
pervasive oscillations almost everywhere, in particular along the
crest of the $|\psi(u,v)|^2$ (Fig.\ 1).
\begin{figure}

\centerline{\begin{tabular}{ccc}
 \epsfig{figure=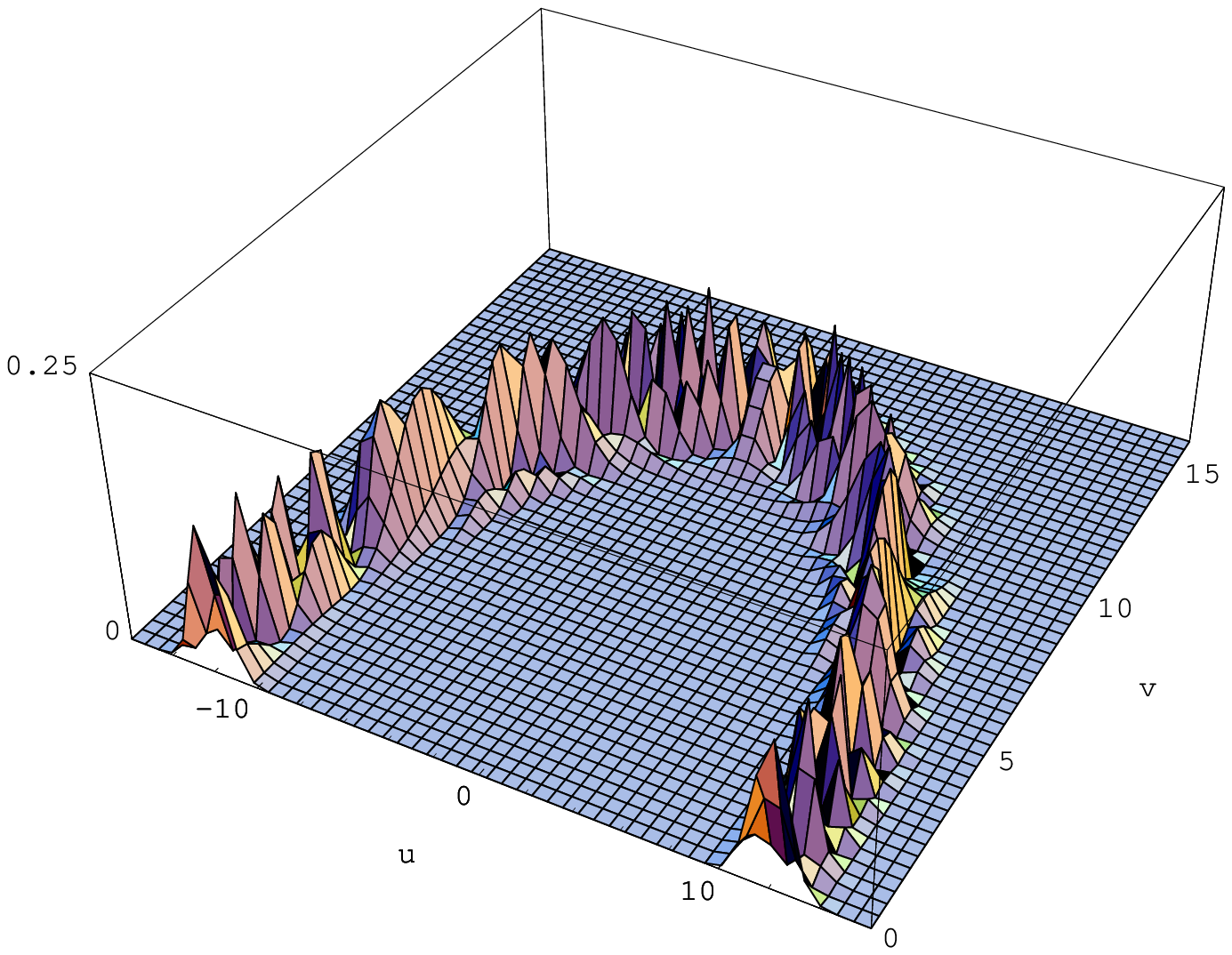,width=8cm}
 &\hspace{2.cm}&
\epsfig{figure=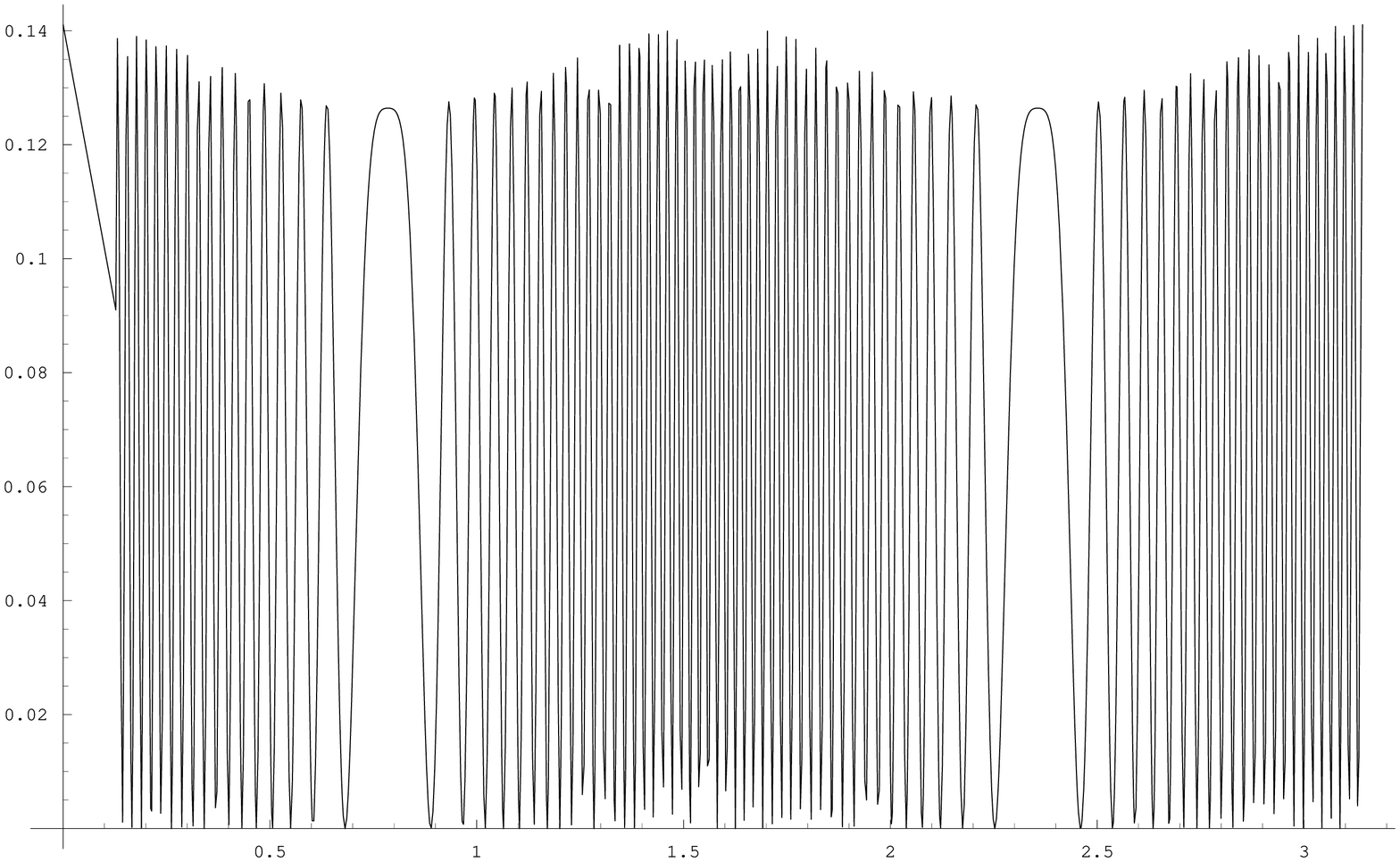,width=6.5cm}
\end{tabular}}
  \caption{Left, plot of the $|\mbox{Re}\,\psi(u,v)|^2$ of the solution to the Wheeler-DeWitt
  Eq.\ (\ref{eqwheeler}) with 130 basis states used in the expansion of the solution. Right, $|\mbox{Re}\,\psi(u,v)|^2$ of this solution along the classical circular
  path \cite{kiefer} (i.e. along its crest .)} \label{oscillation}
\end{figure}
From the Figure it is obvious that the usage of explicit or
implicit FDM would fail in this type of situations. Moreover for
smaller classical path radii, although there will be less
oscillations, the solutions that we attempted to find using FDM
showed divergent behavior at large distances.

We are interested in solving problems which are generalizations of
the one mentioned above.  This gives us motivation to use a
different numerical method to solve this type of problems. This
method, which was first introduced by Galerkin, consists of first
choosing a complete orthonormal set of eigenstates of a, preferably
relevant, hermitian operator to be used as a suitable basis for our
solution. For this numerical method we obviously can not choose the
whole set of the complete basis, as these are usually infinite.
Therefore we make the approximation of representing the solution by
only a finite superposition of the basis functions. By substituting
this approximate  solution into the differential equation, a matrix
equation is obtained. The expansion coefficients of these
approximate  solutions could be determined by eigenvalues and
eigenfunctions of this matrix. This method has been called the
Galerkin Method, and is a subset of the more general Spectral Method
(SM) \cite{SP1,SP2,SP3,maday0}. Spectral methods fall into two broad
categories. The "interpolating", and the "non–interpolating" method.
The first category, which includes the Pseudospectral and the
Spectral Element Methods, divides the configuration space into a set
of grid points. Then one demands that the differential equation be
satisfied exactly at a set of points known as the "collocation" or
"interpolation" points. Presumably, as the residual function is
forced to vanish at an increasingly larger number of discrete
points, it will be smaller and smaller in the gaps between the
collocation points. The "non–interpolating" category includes the
Lanczos tau-method and the Galerkin's method, mentioned above. The
latter is the method that we apply and, in conformity with the usual
nomenclature, we shall simply refer to it as the Spectral Method.
The interesting characteristic of this method is that it is
completely distinct from the usual spatial integration routines,
such as FDM, which concentrate on spatial points. In SM the
concentration is on the basis functions and we expect the final
numerical solution to be approximately independent of the actual
basis used. That is we expect the approximate solution to converge
to the exact solution as the number of basis elements used
increases. This point has been clearly demonstrated by Maday
\textit{et. al.} \cite{maday}. Moreover in this method, the
refinement of the solution is accomplished by choosing a larger set
of basis functions, rather than choosing more grid points, as in the
numerical integration methods. We should note that we are implicitly
assuming that the true solution is expandable in any complete
orthonormal basis such as Fourier, Laguerre\cite{maday2},
Chebyshev\cite{maday3}, or Legendre\cite{maday4} basis. However,
this requirement is usually satisfied for cases of physical
applications.

The paper is organized as follows: In section $2$ and $3$ we
layout the implementation of this method using the quantum
eigenfunctions for the simple harmonic oscillator, henceforth
called the Oscillator basis, and the Fourier basis, respectively.
In section $4$ we solve a particularly interesting example
relevant to quantum cosmology by this method using  both of the
aforementioned basis functions. This problem is a generalization
of the one represented in Eq\@. (\ref{eqwheeler}) and does not
seem to have an exact solution, and we have not been able to solve
this problem by any other numerical method that we tried. In
section $5$ we discuss the accuracy of each of the cases, and make
a comparison between the two. In section $6$ we discuss some
general features of this method.

\section{The Oscillator Basis}
The general PDE equation that we want to solve is a hyperbolic one
cast in the form,
\begin{eqnarray}
H\psi(u,v)=\left\{-\frac{\partial^2}{\partial
u^2}+\frac{\partial^2} {\partial v^2}+ \omega_1^2 u^2-\omega_2^2
v^2+\hat f(u,v)\right\}\psi(u,v)=0, \label{eqgeneralwheeler}
\end{eqnarray}
where $\hat f(u,v)$ is an arbitrary operator of $u$ and $v$, but
with derivatives less than two. Eq.\ (\ref{eqgeneralwheeler}) in
general is not separable, however, any solution can be written as a
superposition of the basis elements,
\begin{eqnarray}
\psi_{m ,n}(u,v)=\alpha_{m}(u)\beta_{n}(v),  \label{eqpsimn}
\end{eqnarray}
where each of the sets $\{\alpha_n\}$ and $\{\beta_m\}$ is an
orthonormal complete set. In this section we take both of them to
be the Oscillator basis \cite{funaro}. That is,
\begin{eqnarray}
\alpha_n(u)=\left(\frac{\omega_1}{\pi}\right)^{1/4}\frac{H_n(
\sqrt{\omega_1}u)} {\sqrt{2^n n!}}e^{-\omega_1 u^2/2},
\label{eqalpha}\\
\beta_n(v)=\left(\frac{\omega_2}{\pi}\right)^{1/4}\frac{H_n(\sqrt{\omega_2}v)
} {\sqrt{2^n n!}}e^{-\omega_2 v^2/2}, \label{eqbeta}
\end{eqnarray}
where $H_n(x)$ denote the Hermite polynomials. In particular the set
$\left\{ \psi_{m ,n}(u,v) \right\} $ is a complete orthonormal set
which can be used to span the zero sector subspace of the Hilbert
space of the hermitian operator $H$, as defined in Eq.\
(\ref{eqgeneralwheeler}). These basis elements are ${\cal L}^2$
measurable square integrable functions on $\bf R^2$ with an inner
product defined in the usual way, so that the orthonormality, for
example, takes the form,
$$ \int \psi_{m,n}(u,v) \psi_{m^\prime ,n^\prime}(u,v) du dv
=\delta_{m,m^\prime} \delta_{n,n^\prime}. $$ We can construct a
general solution as follows,
\begin{eqnarray}
\psi(u,v)=\sum_{m,n} A_{m,n} \alpha_{m}(u) \beta_{n}(v),
\label{eqpsi}
\end{eqnarray}
and can use the following expansion,
\begin{eqnarray}
\hat f(u,v) \psi(u,v)=\sum_{m,n} B_{m,n} \alpha_{m}(u)
\beta_{n}(v),\label{eqf}
\end{eqnarray}
where $B_{m,n}$ are coefficients that can be determined once $\hat
f(u,v)$ is specified. By substituting Eqs.\
(\ref{eqpsi},\ref{eqf}) in Eq.\ (\ref{eqgeneralwheeler}), and
using the differential equation of the Hermite polynomials we
obtain,
\begin{eqnarray}
\sum_{m,n} \Bigl[ [(2m+1)\omega_1 -(2n+1)\omega_2]
A_{m,n}+B_{m,n}\Bigr] \alpha_{m}(u)
\beta_{n}(v)=0.\label{eqmainmatrix}
\end{eqnarray}
Because of the linear independence of $\alpha_{m}(u)$s and
$\beta_{n}(v)$s,
 every term in the summation must satisfy,
\begin{eqnarray}
[(2m+1)\omega_1 -(2n+1)\omega_2] A_{m,n}+B_{m,n}=0.\label{eqAB}
\end{eqnarray}
It only remains to determine the matrix $B$. By using Eqs.
(\ref{eqpsi},\ref{eqf}) we have,
\begin{eqnarray}
\sum_{m,n} B_{m,n} \alpha_{m}(u) \beta_{n}(v)=\sum_{m,n} A_{m,n}
\hat f(u,v) \alpha_{m}(u) \beta_{n}(v).
\end{eqnarray}
By multiply both sides of the above equation by
$\alpha_{m'}(u)\beta_{n'}(v)$ and integrating over the full range
of variables $u$ and $v$, and using orthonormality  of the basis
functions, one finds,
\begin{equation}
B_{m,n}=\sum_{m',n'}\left(
\int_{-\infty}^{\infty}\int_{-\infty}^{\infty} \alpha_{m}(u)
\beta_{n}(v) \hat f(u,v) \alpha_{m'}(u)\beta_{n'}(v) du dv\right
)A_{m',n'}
 \equiv \sum_{m',n'} C_{m,n,m',n'} A_{m',n'}.\label{eqB}
\end{equation}
Therefore we can rewrite Eq.\ (\ref{eqAB}) as,
\begin{eqnarray}
[(2m+1)\omega_1 -(2n+1)\omega_2] A_{m,n}+\sum_{m',n'}
C_{m,n,m',n'}\,\, A_{m',n'}=0. \label{eqAC}
\end{eqnarray}
It is obvious that the presence of the operator $\hat f(u,v)$ in
Eq.\ (\ref{eqgeneralwheeler}), leads to nonzero coefficients
$C_{m,n,m',n'}$ in Eq.\ (\ref{eqAC}), which in principle could
couple all of the matrix elements of $A$. For the usual choices of
$\hat f(u,v)$, {\em e.g.\ }the choice presented in
\cite{wavepacket}: $\hat f=\frac{9}{4}k (u^2-v^2)^{1/3}$, the
problem, to the best of our knowledge, is not analytically
solvable. Therefore we have to resort to a numerical solution. In
general the number of basis elements are at least countably
infinite. The aforementioned coupling of terms in the main matrix
Eq.\ (\ref{eqAC}) forces us to make the approximation of using a
finite basis. It is easy to see that the more basis functions we
include, the closer our solution will be to the exact one. We
select the first $N$ basis functions in each direction, that is
$m$ and $n$ run from $1$ to $N$. Then we replace the square matrix
$A$ with a column vector $A'$ with $N^2$ elements, so that any
element of $A$ corresponds to one element of $A'$. With this
replacement, Eq\. (\ref{eqAC}) can be written as,
\begin{eqnarray}
D\, A'=0, \label{eqmatrix}
\end{eqnarray}
where $D$ is a square matrix with $N^2 \times N^2$ elements which
can be obtained from Eq.\ (\ref{eqAC}). Looked upon as an eigenvalue
equation, {\it i.e.} $DA'_a=a A'_a$, the matrix $D$ has $N^2$
eigenvectors. However, for constructing the acceptable
wavefunctions, {\em i.e. }the ones satisfying the WD Eq.\
(\ref{eqgeneralwheeler}), we only require eigenvectors which span
the null space of the matrix $D$. That is, due to Eq.\ (\ref{eqAC})
we will have exactly $N$ null eigenvectors which will be linear
combination of our original eigenfunctions introduced in Eq.\
(\ref{eqpsimn}). After finding these $N$ eigenvectors $A'^i$ ($
i=1,2,3,...,N$), we can find the corresponding elements of the
matrix $A$, $A_{m,n}^{i}$. Therefore the wavefunction can be
expanded as,
\begin{eqnarray}
\psi(u,v)=\sum_{i} \lambda^i \psi^i(u,v), \label{eqpsifinal}
\end{eqnarray}
where $\psi^i(u,v)$ are the appropriate eigenfunctions for the
problem ({\it i.e.} Eq.\ (\ref{eqgeneralwheeler})),
\begin{eqnarray}
\psi^i(u,v)=\sum_{m,n} A_{m,n}^i \alpha_m(u) \beta_n(v).
\label{eqpsii}
\end{eqnarray}
In Eq.\ (\ref{eqpsifinal}) $\lambda^i$s are arbitrary complex
constants to be determined by the initial conditions.

\section{The Fourier Basis}
As mentioned before, any complete orthonormal set can be used for
this SM. In this section we use the Fourier series basis as a
second example. That is, since we need to choose a finite subspace
of a countably infinite basis, we restrict ourselves to a finite
square region of sides $2L$. This means that we can expand the
solution as,
\begin{eqnarray}
\psi(u,v)=\sum_{i,j=1}^2 \sum_{m,n} A_{m,n,i,j} \,\,\,
g_i\left(\frac{m \pi u}{L}\right)\,\,\, g_j\left(\frac{n \pi
v}{L}\right), \label{eqpsitrigonometric}
\end{eqnarray}
where,
$$
\left\{
  \begin{array}{ll}
    g_1\left(\frac{m \pi
u}{L}\right)=\sqrt{\frac{2}{R_{m} L}}\sin\left(\frac{m \pi
u}{L}\right), &  \nonumber \\
   g_2\left(\frac{m \pi
u}{L}\right)=\sqrt{\frac{2}{R_{m} L}}\cos\left(\frac{m \pi
u}{L}\right). & \nonumber \\
  \end{array}\nonumber
\right. \mbox{and}\,\,\, R_{m} =\left\{ \begin{array}{ll}
    1, & m\ne0 \nonumber \\
    2, & m=0 \nonumber \\
\end{array}
\right.
$$
That is in the Fourier basis we assume periodic boundary
condition. By referring to the WD Eq.\ (\ref{eqgeneralwheeler}), we
realize that in the Fourier basis it is appropriate to introduce
$\hat f'$ as,
\begin{eqnarray}
\hat f'(u,v)=f(u,v)+\omega_1^2 u^2-\omega_2^2 v^2,\label{eqf'}
\end{eqnarray}
and, in analogy with Eq.\ (\ref{eqf}), we can make the following
expansion,
\begin{eqnarray}
\hat f'(u,v) \psi(u,v)=\sum_{i,j} \sum_{m,n} B'_{m,n,i,j} \,\,\,
g_i\left(\frac{m \pi u}{L}\right)\,\,\, g_j\left(\frac{n \pi
v}{L}\right).
\end{eqnarray}
By following steps analogous to those of Eqs.
(\ref{eqmainmatrix}-\ref{eqB}) we obtain,
\begin{eqnarray}
\left[\left(\frac{m \pi }{L}\right)^2 -\left(\frac{n \pi
}{L}\right)^2\right] A_{m,n,i,j}+B'_{m,n,i,j}=0,\label{eqAB'}
\end{eqnarray}
where
\begin{eqnarray}
B'_{m,n,i,j}\hspace{-3mm}&=& \sum_{m',n',i',j'} \left[
\int\hspace{-2mm}\int_{-L}^{L} g_{i}\left(\frac{m \pi u}{L}\right)
g_{j}\left(\frac{n \pi v}{L}\right) \hat f'(u,v)
g_{i'}\left(\frac{m' \pi u}{L}\right) g_{j'}\left(\frac{n' \pi
v}{L}\right)du dv\right]A_{m',n',i',j'}\nonumber\\ &=&
\sum_{m',n',i',j'} C'_{m,n,i,j,m',n',i',j'}\,\,
A_{m',n',i',j'},\label{B'}
\end{eqnarray}
Therefore we can rewrite Eq.\ (\ref{eqAB'}) as
\begin{eqnarray}
\left[(m \pi )^2 -(n \pi )^2\right] A_{m,n,i,j}+
\sum_{m',n',i',j'} C'_{m,n,i,j,m',n',i',j'}\,\,
A_{m',n',i',j'}=0.\label{eqAC'}
\end{eqnarray}
In this case, we select $4N^2$ basis functions. Using reasoning
analogous to the previous case, for example by defining the column
vector $A'$ out of the matrix $A$, we transform Eq.\ (\ref{eqAC'})
to
\begin{eqnarray}
D'\, A'=0. \label{eqmatrix2}
\end{eqnarray}
Where, as before, $D'$ is a square matrix now with $(2N)^2 \times
(2N)^2$ elements which can be easily obtained from Eq.\
(\ref{eqAC'}). After finding the $4N^2$ eigenvectors of $D'$, we
select the 2N ones with zero eigenvalue, {\em i.e.} $A'^k$ ($
k=1,2,3,...,2N$). We can then find the corresponding elements of
matrix $A$, $A^k_{m,n,i,j}$. Therefore, the wavefunction can be
expanded as
\begin{eqnarray}
\psi(u,v)= \sum_{k} \lambda^k \psi^k(u,v), \label{eqpsifinal2}
\end{eqnarray}
where, as before, $\psi^k(u,v)$ are the appropriate eigenfunctions
for the problem ({\it i.e.} Eq.\ (\ref{eqgeneralwheeler})),
\begin{eqnarray}
\psi^k(u,v)=\sum_{m,n,i,j} A^k_{m,n,i,j}\,\,\,  g_i\left(\frac{m
\pi u}{L}\right)\,\,\, g_j\left(\frac{n \pi v}{L}\right).
\label{eqpsii2}
\end{eqnarray}
Here $\lambda^k\,\,$s in Eq.\ (\ref{eqpsifinal2}) are again
arbitrary complex constants to be determined by the initial
conditions.

Now we apply this method to one of the examples stated above which
happens to be relevant in quantum cosmology, and was our original
motivation for using this method.

\section{Application of the Spectral Method to a Specific Example}
\begin{figure}
\centerline{\begin{tabular}{ccc} \epsfig{figure=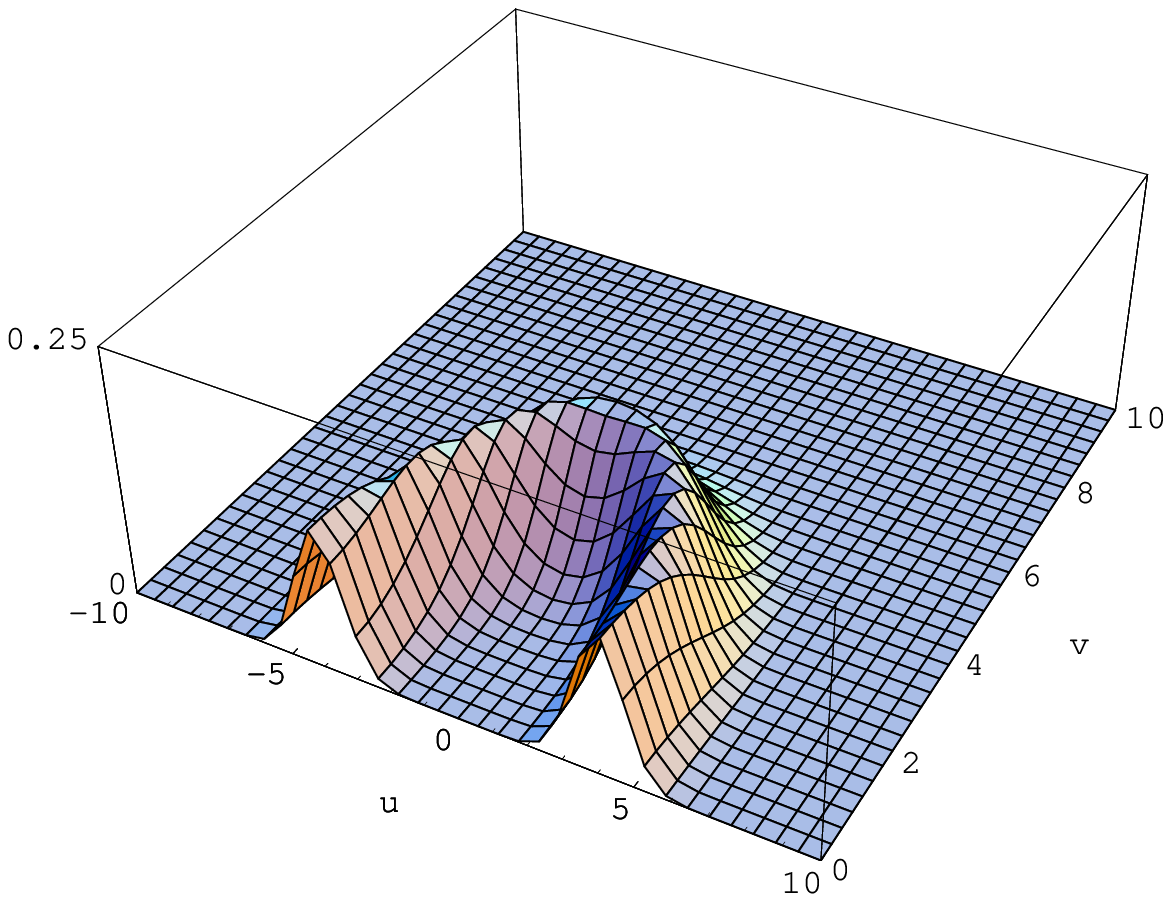,width=8cm}
 &\hspace{2.cm}&
\epsfig{figure=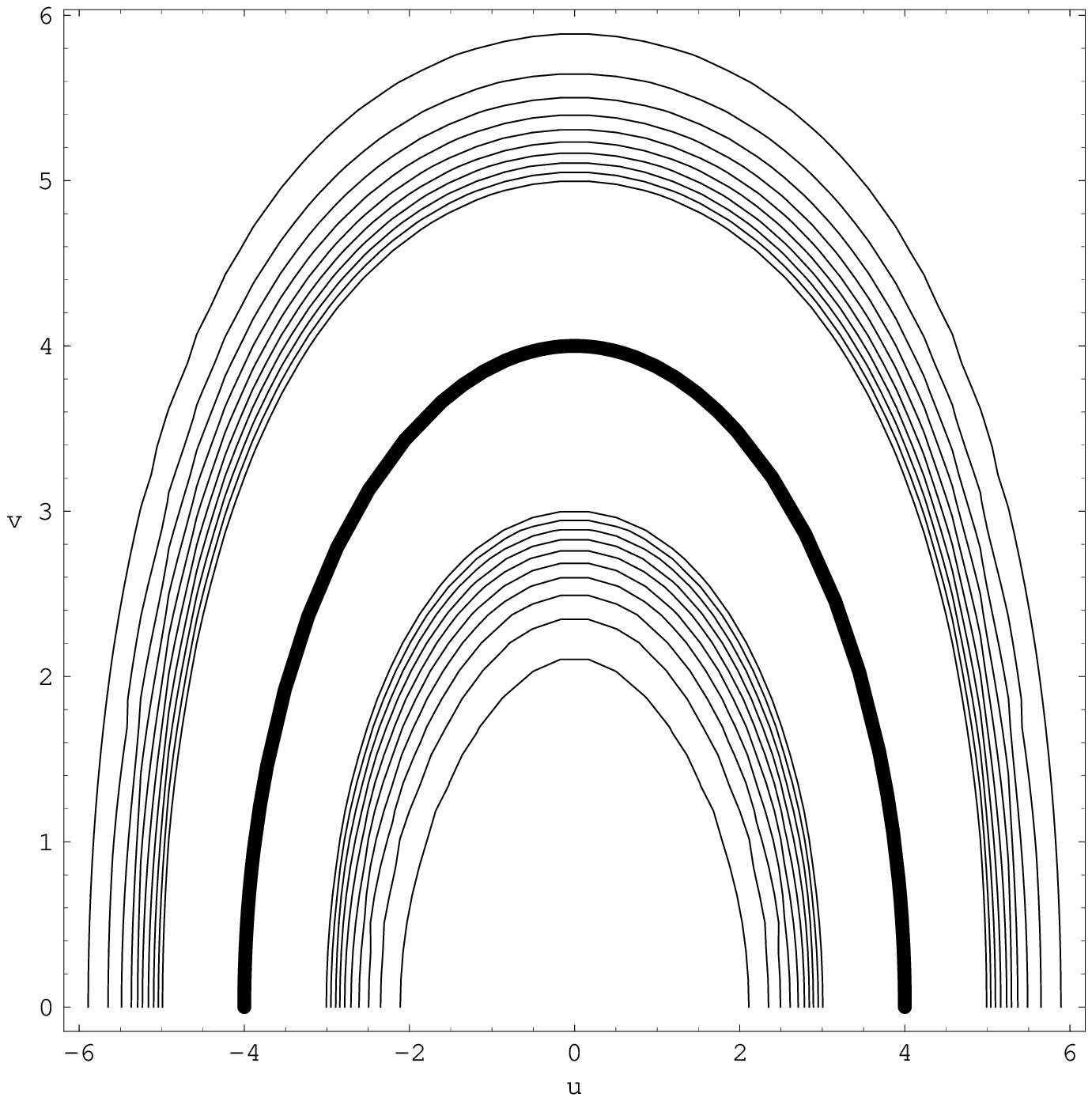,width=6.5cm}
\end{tabular}}
\caption{Oscillator Basis: Left, the absolute value squared of the
wave packet $| \psi(u,v)|^2$ for $\chi=4$ and $N=35$, $k=0$.
Right, the contour plot of the same figure with the classical path
superimposed as the thick solid line.} \label{fig1}
\end{figure}

For a specific example, we consider a hyperbolic PDE which happens
to be the Wheeler-DeWitt equation for the Robertson-Walker quantum
cosmology with non-zero curvature,
\begin{eqnarray}
\left\{-\frac{\partial^2}{\partial u^2}+\frac{\partial^2}
{\partial v^2}+ \omega^2_1 u^2- \omega_2^2 v^2+\frac{9}{4}k
(u^2-v^2)^{1/3}\right\}\psi(u,v)=0. \label{eqwheeler2}
\end{eqnarray}
As mentioned before, the case $k=0$ is exactly solvable
\cite{wavepacket} and has a closed form solution in the Oscillator
basis. We shall state these solutions here for illustrative
purposes, especially for motivating the choice of initial
conditions and comparison with the non-trivial cases, {\it i.e}
$k\ne0$, which shall be solved by this method next. In this case
$f(u,v)=0$ in Eq.\ (\ref{eqgeneralwheeler}), so the matrix $B$ in
Eq.\ (\ref{eqAB}) is also zero due to Eq.\ (\ref{eqf}). Therefore,
Eq.\ (\ref{eqAB}) reduces to,
\begin{eqnarray}
\left[(2m+1)\omega_1-(2n+1)\omega_2\right] A_{m,n}=0. \label{eq19}
\end{eqnarray}
This means $A_{m,n}=0$ for,
\begin{eqnarray}
\omega_2\neq\frac{2m+1}{2n+1}\omega_1.
\end{eqnarray}
That is we have nontrivial solutions only for $\omega_1$ and
$\omega_2$ being a rational multiple of each other. Choosing
$\omega_1=\omega_2$ for simplicity, we have $n=m$ and the
expansion of the wavefunction (Eq.\ (\ref{eqpsi})) reduces to,
\begin{eqnarray}
\psi(u,v)=\sum_{m} A_{m} \alpha_{m}(u) \beta_{m}(v), \label{eq21}
\end{eqnarray}
where $A_{m} \equiv A_{m,m}$ are complex expansion coefficients
that can be determined by applying initial conditions ({\it i.e.
}specifying $\psi(u,0)$ and $\frac{\partial\psi}{\partial
v}|_{v=0}$). In references \cite{wavepacket,kiefer} the authors
considered the following initial condition on the wave function,
\begin{equation}
\psi(u,0)=\frac{1}{2
\pi^{1/4}}\left(e^{-(x-\chi)^2/2}+e^{-(x+\chi)^2/2}\right).\label{eqpsi0}
\end{equation}
We can decompose any initial wavefunction in the Oscillator basis.
For the choice presented in Eq. (\ref{eqpsi0}) we have,
\begin{eqnarray}
\psi(u,0)=\sum_{n}^{'} c_{n} \alpha_{n}(u),\hspace{5mm}
\mbox{where} \hspace{5mm}c_n=e^{-\frac{1}{4}
|\chi|^2}\frac{\chi^n}{\sqrt{2^n n!}}\, , \label{eqcondition1}
\end{eqnarray}
and $\chi$ is an arbitrary complex number. The prime on the
summation denotes the restriction that $n$ is even. These
coefficients are same as those of the coherent states of a one
dimensional simple harmonic oscillator. With this choice of
coefficients we expect that a classical-quantum correspondence
should be manifest. The canonical choice for initial slope is
\cite{wavepacket},
\begin{eqnarray}
\left. \frac{\partial\psi}{\partial v}\right|_{v=0}=\sum_{n}
\frac{c_{n} \alpha_{n}(u)H'_n(0)}{\frac{(-1)^{(n/2)} n!}{(n/2)!}},
\label{eqcondition2}
\end{eqnarray}
where $H_n$s are the Hermite polynomials and prime denotes
differentiate respect to $v$. By choosing $\chi$ to be a real, the
classical paths corresponding to these solutions can be shown to
be circles with radii $\chi$. We have found that $35$ basis
functions are sufficient for finding the solution to an accuracy
of about $10^{-8}$ , when the classical radius of the wave packet
($\chi$) is less than $4$ (Fig.\ \ref{fig1}). As can be seen in
the figure, and also for all the cases presented in
\cite{wavepacket}, the classical-quantum correspondence is
manifest. Note that the wave packet has compact support and in the
oscillator basis, the truncation of the basis functions
automatically restricts the solution to have this property. We
only need to choose the configuration space domain to be large
enough.

At this point it seems to us that a brief mention of the relevant
dynamical equations for the classical cosmology might be helpful,
at least for completeness, especially in light of the fact that
``classical-quantum correspondence" that we keep mentioning in
this paper is particularly important in physics. Moreover, the
non-linearity and the moving singular behavior of these equations
would become apparent. These equations are (see, for example,
\cite{siamak}),
\begin{eqnarray}
\ddot{u}+u+\frac{3k}{2}\frac{u}{(u^2-v^2)^{2/3}}=0,\label{u''}
\end{eqnarray}
\begin{eqnarray}
\ddot{v}+v+\frac{3k}{2}\frac{v}{(u^2-v^2)^{2/3}}=0,\label{v''}
\end{eqnarray}
\begin{eqnarray}
\dot{u}^2+u^2-\dot{v}^2-v^2+\frac{9}{4}k(u^2-v^2)^{1/3}=0.\label{u2}
\end{eqnarray}
Here the $u$ and $v$ variables are functions of time. Equations
(\ref{u''}) and (\ref{v''}) are the dynamical equations and, Eq.\
(\ref{u2}) is the zero energy constraint, from which the
Wheeler-deWitt Eq.\ (\ref{eqwheeler2}) arises. An appropriate
initial conditions is the following,
\begin{eqnarray}
 u(0)=-\chi,\hspace{1cm} v(0)=0,\hspace{1cm}
\dot{u}(0)=0,\hspace{1cm}
\dot{v}(0)=\dot{v}_0,\label{eqclassicalic}
\end{eqnarray}
where  $\chi$  can be treated as a free parameter and $\dot{v}_0$
is adjusted so that Eq.\ (\ref{u2}) is satisfied. With this choice
of initial conditions the classical path would be, as mentioned
before, exactly a circle in the $k=0$ case, as depicted in Fig.\
\ref{fig1}. In the general case of $\omega_1\ne \omega_2$, these
solutions would give Lissajous figures.

For the case $k\ne 0$, the problem is not exactly solvable in
quantum cosmology and we will use the SM to get an approximate
solution. We have to mention that the corresponding equations for
classical cosmology, Eqs. (\ref{u''})-(\ref{u2}), are a set of
non-linear, coupled ODEs with moving singularities which are not
exactly solvable either. However, a general method for solving them
has been presented in \cite{siamak2}, and this is the method we
shall use. Also a detailed explanation of the physical setting of
the problem in the classical domain has been presented in
\cite{siamak}. On one hand, we need to choose a set of appropriate
initial conditions for both the classical and quantum cases which
would make their correspondence manifest when they are superimposed,
as in Fig.\ \ref{fig1} which was for the $k=0$ case. However, we
should note  that the values of $\dot{v}_0$ in the classical case
depends on $k$. On the other hand, an appropriate choice for the
initial conditions should be such that we could easily compare our
results with the $k=0$ case. Therefore, here we choose the same
initial conditions for the quantum cosmology Eqs.\
(\ref{eqcondition1},\ref{eqcondition2}), and classical cosmology
Eq.\ (\ref{eqclassicalic}), as for the $k=0$ case. An alternative
would be to choose appropriate initial canonical slope for the
quantum cosmology case \cite{wavepacket2}. Having set the initial
conditions, we proceed to solve the quantum cosmology problem for
the case $k\ne0$ with the Oscillator basis.

\begin{figure}
\centerline{\begin{tabular}{ccc} \epsfig{figure=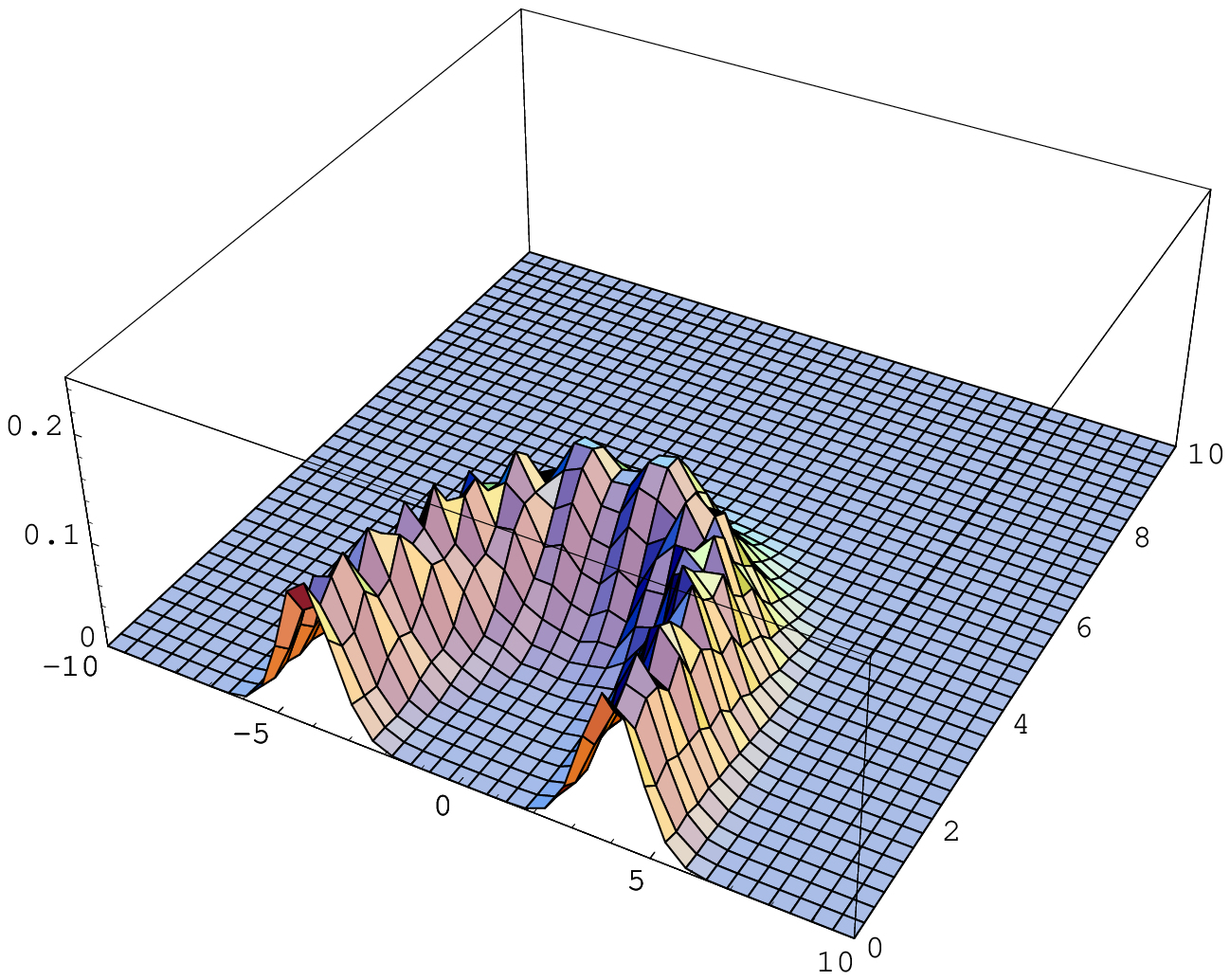,width=8cm}
 &\hspace{2.cm}&
\epsfig{figure=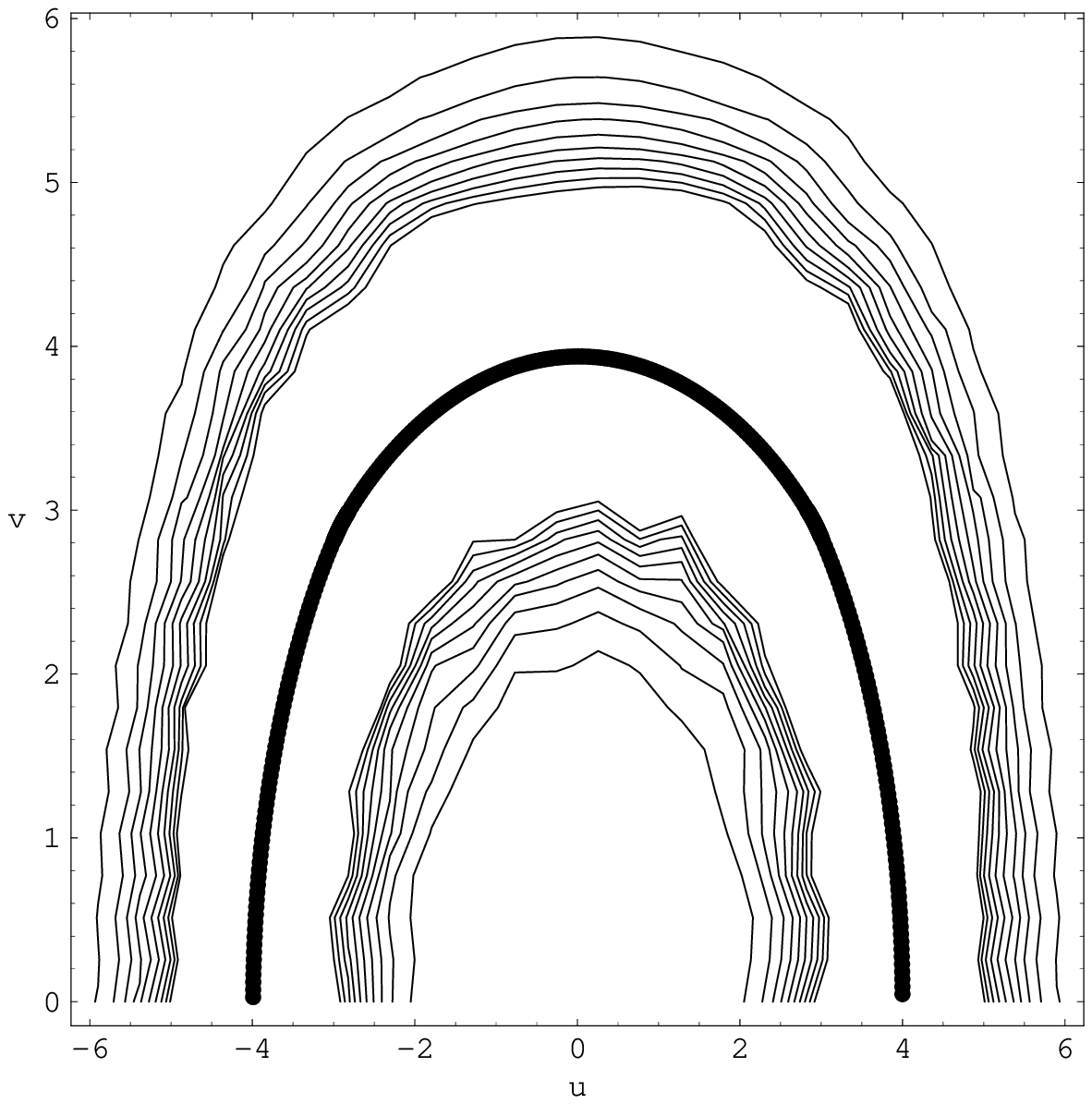,width=6.5cm}
\end{tabular}}
\caption{Oscillator Basis: Left, the absolute value squared of the
wave packet $| \psi(u,v)|^2$ for $\chi=4$ and $N=35$, $k=+1$.
Right, the contour plot of the same figure with the classical path
superimposed as the thick solid line.} \label{fig2}
\end{figure}

Both the classical and quantum solutions for the case $k=+1$ are
shown in Fig.\ \ref{fig2}. We can see that the general behavior of
this case is similar to the $k=0$ case. In particular, again we
have a very good classical-quantum correspondence. However,
although the extreme points of the solution do not change as
compared to the $k=0$ case, the whole pattern is a little wider.
Moreover, the solution ($|\psi|^2$) is not as smooth as $k=0$
case.
\begin{figure}
\centerline{\begin{tabular}{ccc} \epsfig{figure=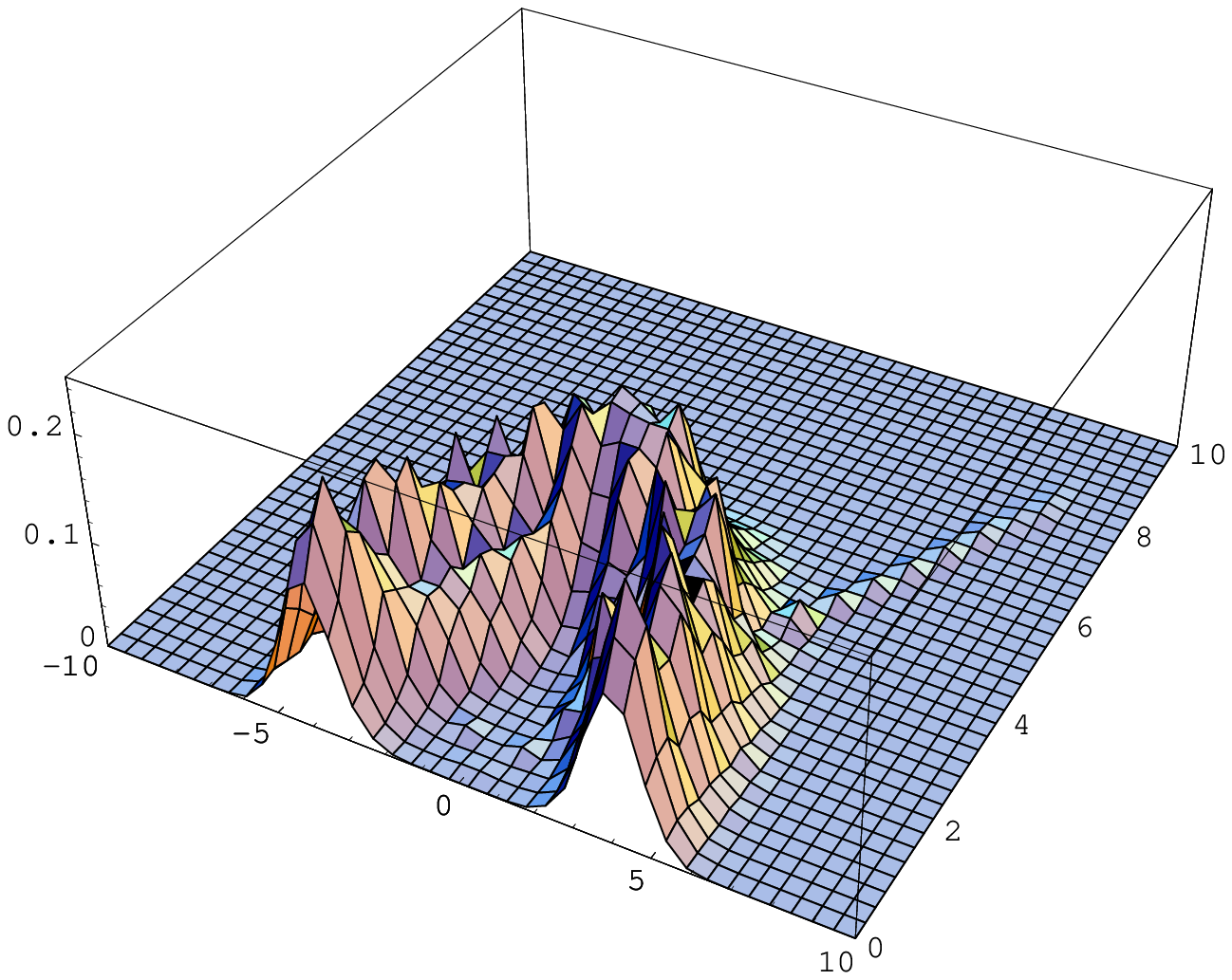,width=8cm}
 &\hspace{2.cm}&
\epsfig{figure=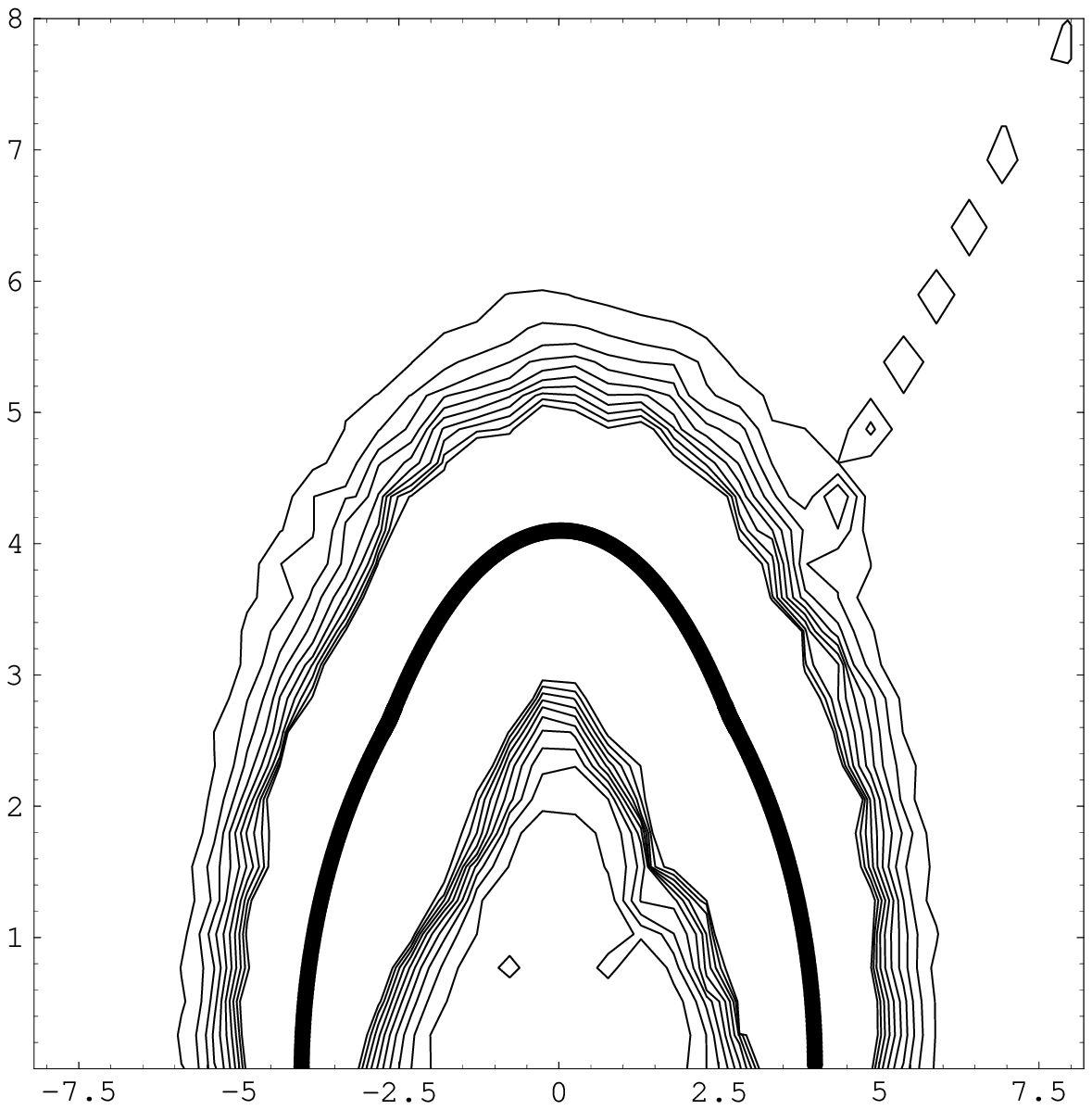,width=6.5cm}
\end{tabular}}
\caption{Oscillator Basis: Left, the absolute value squared of the
wave packet $| \psi(u,v)|^2$ for $\chi=4$ and $N=35$, $k=-1$.
Right, the contour plot of the same figure with the classical path
superimposed as the thick solid line.} \label{fig3}
\end{figure}

Both the classical and quantum solutions for the case $k=-1$ are
shown in Fig.\ \ref{fig3}. We can see that the general behavior of
this case is also similar to the $k=0$ case. In particular, again
we have a very good classical-quantum correspondence. However,
although the extreme points of the solution do not change as
compared to the $k=0$ case, the whole pattern is a little
narrower. Moreover, the solution ($|\psi|^2$) is not as smooth as
$k=0$ case. More importantly, when $k=-1$ we encounter a new
characteristic of the solution. Note that the $u^2-v^2$ term in
Eq.\ (\ref{eqwheeler2}) is usually dominant and causes stability
of the solutions, for all $k$. However, the $\frac{9}{4}
k(u^2-v^2)^{(1/3)}$ term in that equation could cause instability
near $u=\pm v$ lines, only in the case of $k<0$. However this is
precisely where the dominant term vanishes. Therefore we do expect
numerical instabilities along these two lines for this case. These
instabilities can be seen in Fig.\ \ref{fig3}. However, due to
numerical approximations made, the instabilities seem to be more
pronounced along the line $u=v$ in our solution. We have not been
able to the pinpoint the exact source of this numerical asymmetry
in the instability.

\begin{figure}
\centerline{\begin{tabular}{ccc} \epsfig{figure=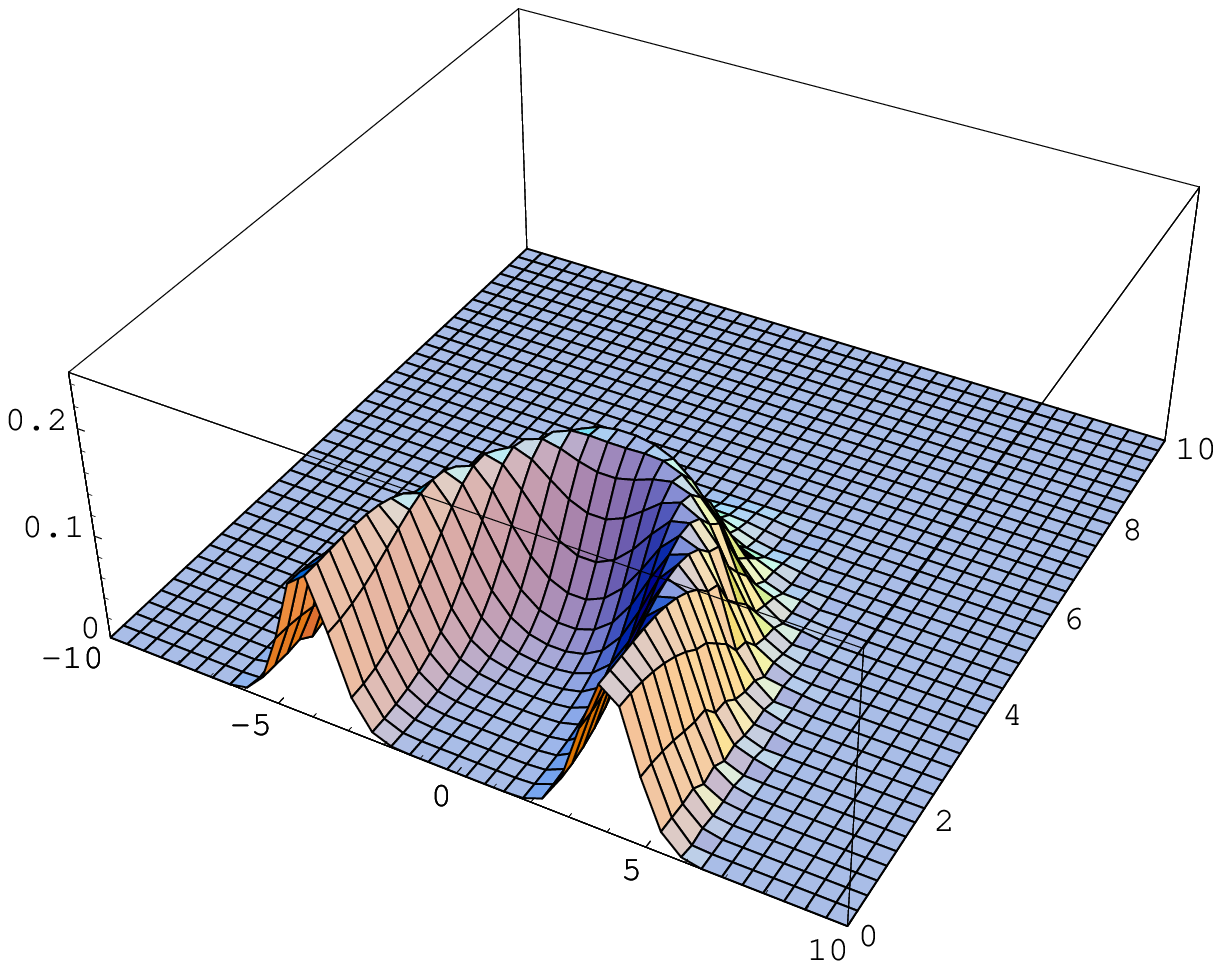,width=8cm}
 &\hspace{2.cm}&
\epsfig{figure=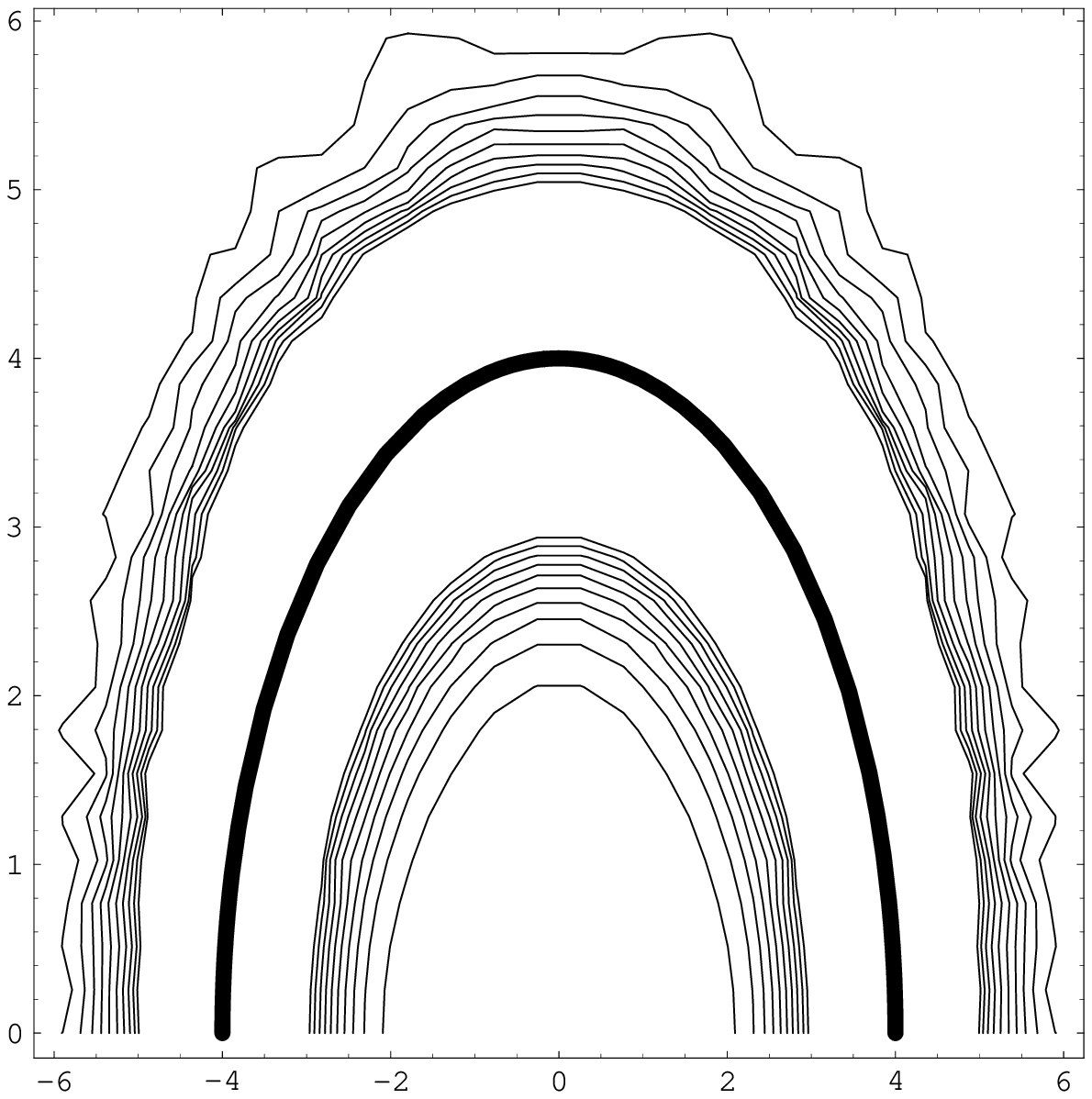,width=6.5cm}
\end{tabular}}
\caption{Fourier Basis: Left, the absolute value squared of the
wave packet $|\psi(u,v)|^2$ for $\chi=4$ and $N=35$, $k=0$. Right,
the contour plot of the same figure with the classical path
superimposed as the thick solid line.} \label{fig4}
\end{figure}

\begin{figure}
\centerline{\begin{tabular}{ccc} \epsfig{figure=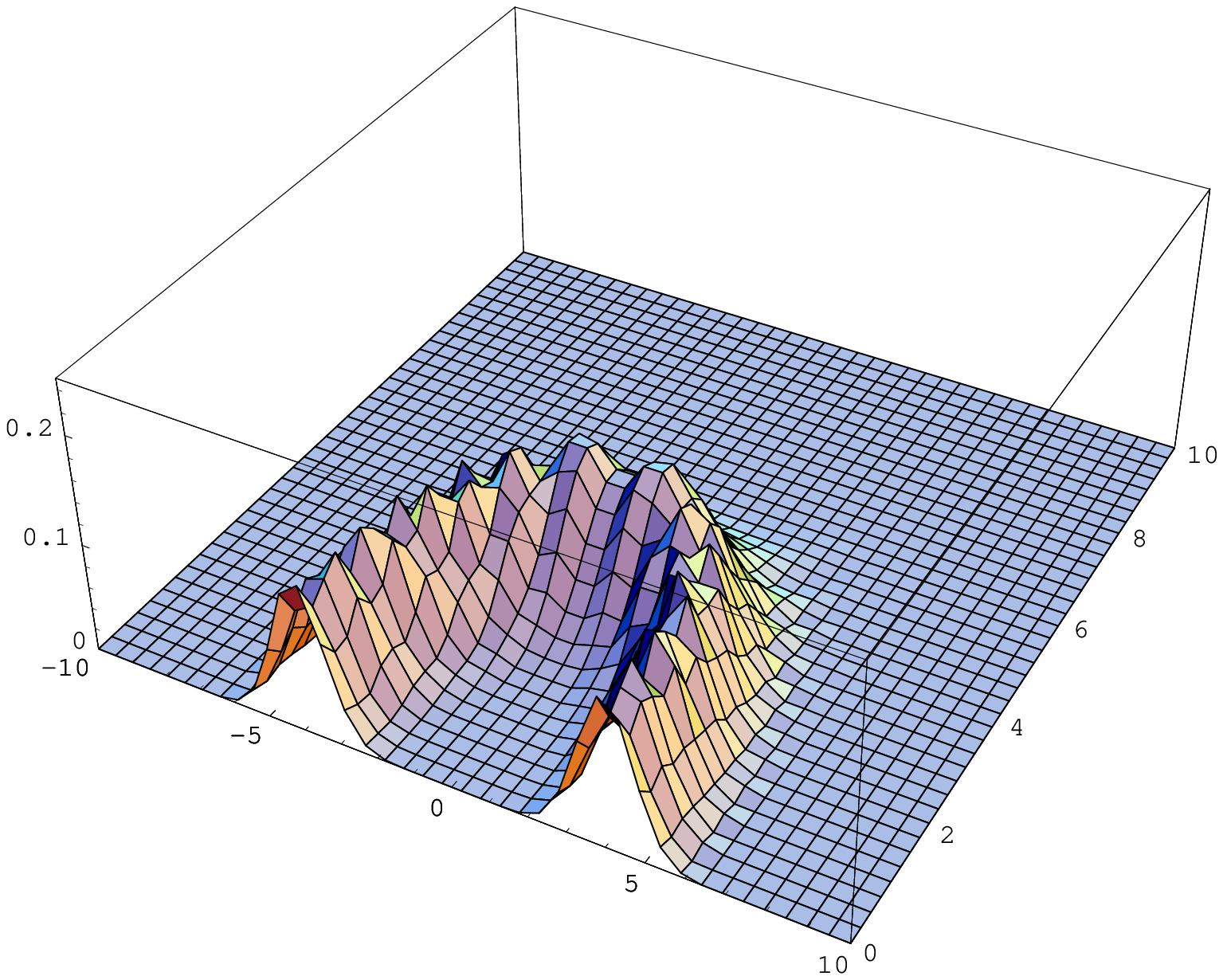,width=8cm}
 &\hspace{2.cm}&
\epsfig{figure=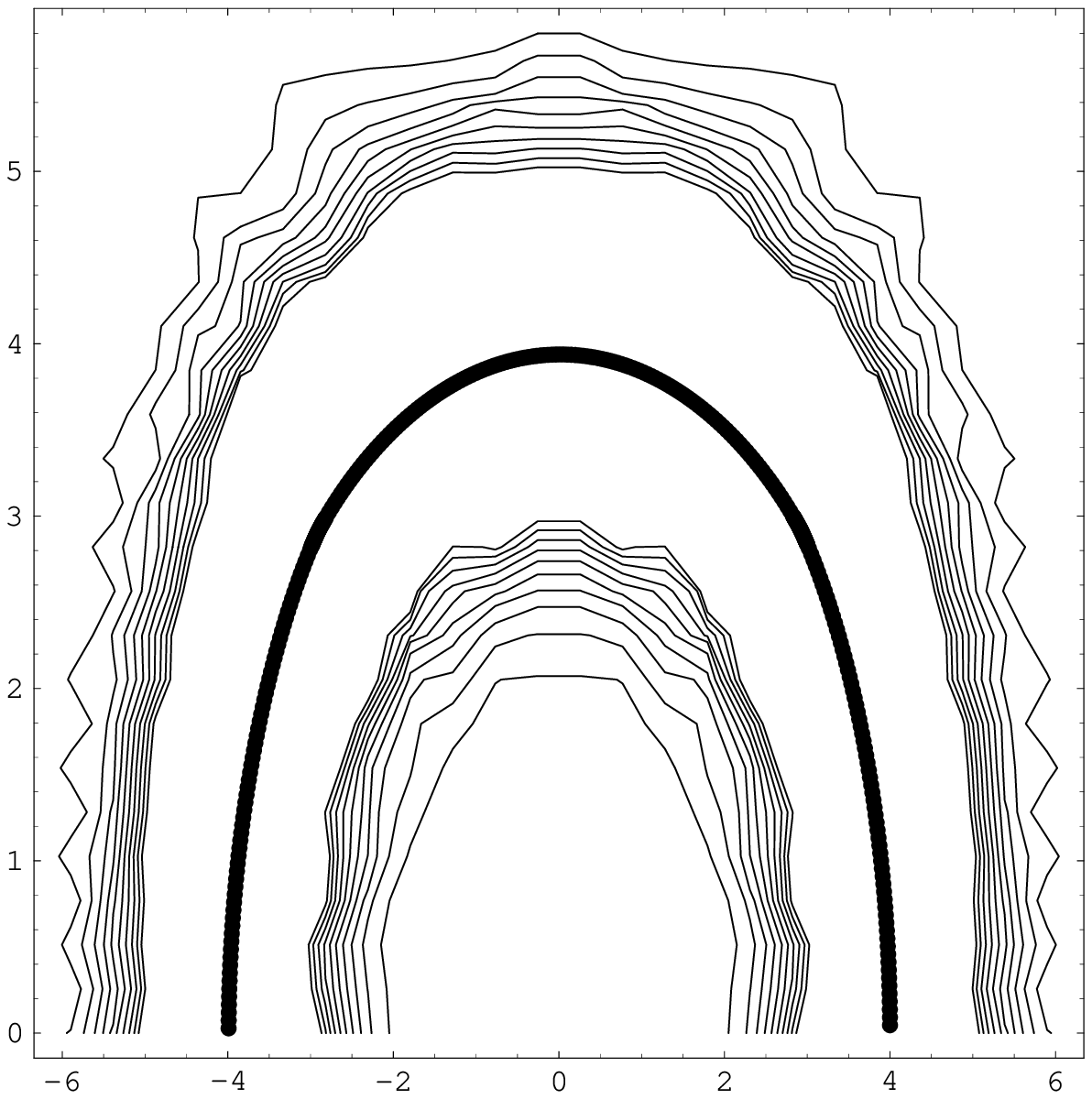,width=6.5cm}
\end{tabular}}
\caption{Fourier Basis: Left, the absolute value squared of the
wave packet $| \psi(u,v)|^2$ for $\chi=4$ and $N=35$, $k=1$.
Right, the contour plot of the same figure with the classical path
superimposed as the thick solid line.} \label{fig5}
\end{figure}

\begin{figure}
\centerline{\begin{tabular}{ccc} \epsfig{figure=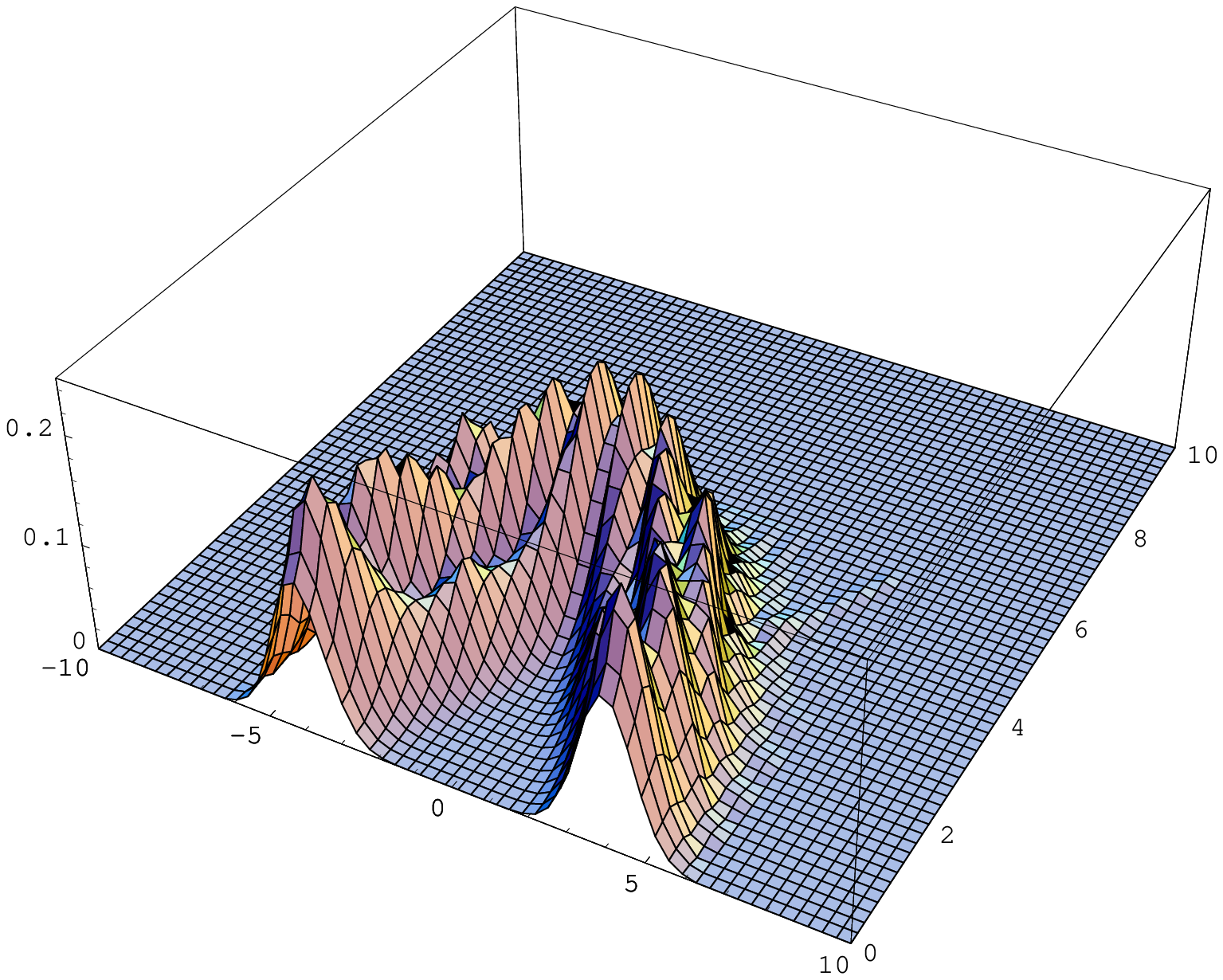,width=8cm}
 &\hspace{2.cm}&
\epsfig{figure=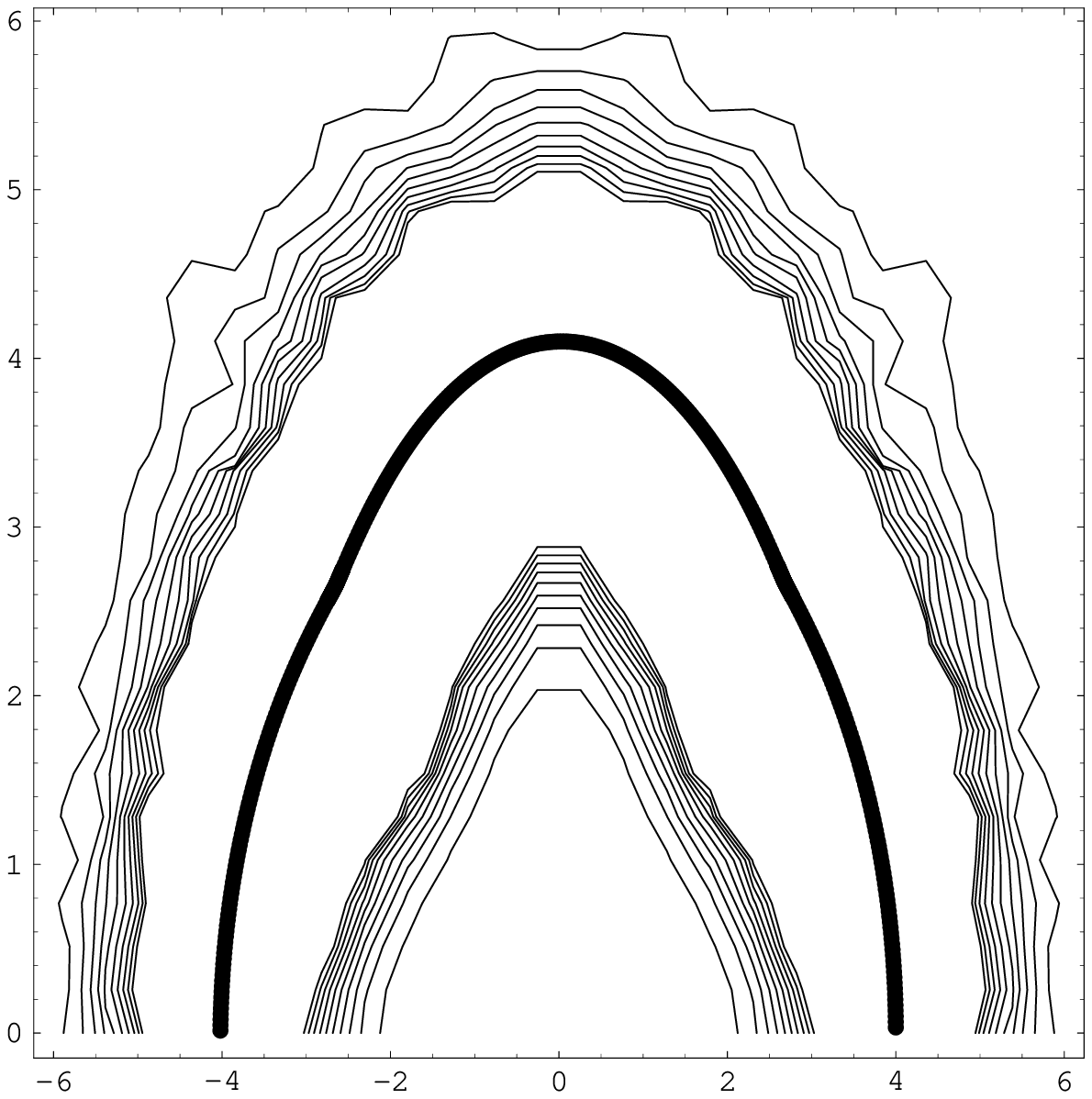,width=6.5cm}
\end{tabular}}
\caption{Fourier Basis: Left, the absolute value squared of the
wave packet $| \psi(u,v)|^2$ for $\chi=4$ and $N=35$, $k=-1$.
Right, the contour plot of the same figure with the classical path
superimposed as the thick solid line.} \label{fig6}
\end{figure}

It would be an interesting comparison to solve exactly the same
problem in the Fourier basis. By using the procedure mentioned in
section 3, and again choosing exactly $35$ basis functions for
ease of comparison, we easily find the solutions for $k=0$, $k=1$
and $k=-1$, which are shown in Figs. (\ref{fig4}-\ref{fig6}),
respectively.

\section{Comparison of the Oscillator and Fourier Bases}
Here, we compare the results of the Spectral Method using the two
finite bases. To this end, we first discuss the errors of the
solutions. In the case $k=0$ we have the exact solutions as an
infinite series. For small radii, {\em e.g.\ } $\chi=4$, the
difference between the exact solution and a series solution with
$120$ oscillator terms is absolutely negligible. Therefore, we can
safely substitute this finite series for the exact one. For
computing errors, we divide the $2D$ base domain into $M^2$ grid
points. Then, we average the square of the absolute value of the
difference of the exact solution with that obtained by the $35$
Fourier or oscillator basis functions on the grid points:
\begin{equation}
\delta^2_{k=0}=\frac{\sum_{i,j}^M
|\psi_{35}(i,j)-\psi_{120}(i,j)|^2}{\sum_{i,j}^M
|\psi_{120}(i,j)|^2} .
\end{equation}
In the case $k\ne0$ the problem is not exactly solvable, so for
comparison purposes we define a measure for the error as the
average square of the absolute value of the difference of the
solutions with, for example, $35$ and $30$ basis functions. That
is,
\begin{equation}
\delta^2_{k\ne0}=\frac{\sum_{i,j}^M
|\psi_{35}(i,j)-\psi_{30}(i,j)|^2}{\sum_{i,j}^M
|\psi_{35}(i,j)|^2} .
\end{equation}
\begin{center}
\begin{tabular}[width=1c]{|c|c|c|}
\hline & $\delta_{Fourier}$& $\delta_{oscillator}$\\
\hline

$k=0$ & $6.70875 \times 10^{-2}$  & $4.16459 \times10^{-7}$   \\
\hline
$k=+1$ &$4.01423 \times10^{-3}$    & $3.84836 \times10^{-2}$   \\
\hline
$k=-1$ & $2.83746 \times10^{-3}$   & $3.64503 \times10^{-2}$   \\
\hline
\end{tabular}

\end{center}
\begin{center}
Table I. Errors for the Oscillator and Fourier basis
\end{center}
From Table I we can conclude that the oscillator basis is more
appropriate than Fourier basis for the case $k=0$. It seems that
this is only due to the fact that the Oscillator basis is the exact
solution of the problem in this case. However, the Fourier basis is
more appropriate for the case $k\ne0$. The main reason for this
seems to us to be the fact that the solutions have compact support,
so we have an extra parameter that we can adjust (the size of
spatial domain (2L)) and this yields better results. In fact one can
use this freedom to set up an optimization procedure \cite{VISM}.
\section{Discussion}
Here we have exhibited the implementation of the SM (Galerkin) for
solving hyperbolic PDEs. We use finite basis of Oscillator or
Fourier eigenfunctions, for example, and show that in some cases
where the popular numerical methods such as FDM or FEM fail, this
method gives reasonable results very easily. The requirement that
the solution should be expandable in a complete orthonormal basis is
crucial. However, certain bases might be more appropriate for a
given problem. For example, if the wave packets of previous section
was not damped strongly in $u$ and $v$ directions, choosing
oscillator basis might not have been as appropriate. We have found,
much to our surprise, that the main source of error is the numerical
integrations, as compared to varying the number of basis elements.
This arise due to the fact that in order to find the coefficients
$C(m,n,m',n')$, we need to calculate $N^4$ two fold integrations,
where $N$ denotes the number of basis elements. This is the most
time consuming part of the procedure. Using programs with refined
integration routines such as Mathematica would have consumed too
much time even at their default levels. So we used the simple
trapezoid integration technique using Fortran, and this severely
limited our accuracy when we refrained from refining our mesh
excessively to avoid spending too much time. However, calculations
of $C(m,n,m',n')$s can be parallelized because these coefficients
are independent. In some cases we can calculate some of the
integrals analytically ({\em e.g.\ } the values of
$C'_{m,n,i,j,m',n',i',j'}$ in Eq.\ (\ref{B'}) when $k=0$), which
decreases the computation time. The memory consuming part of the
procedure is solving the matrix equation (Eq.\ (\ref{eqmatrix})) and
this increases proportional to $N^4$.

\vskip20pt\noindent {\large {\bf
Acknowledgement}}\vskip5pt\noindent This research has been
supported by the office of research of Shahid Beheshti University
under Grant No. 500/3787. 
\end{document}